# Data-Driven Response Regime Exploration and Identification for Dynamical Systems

Dr. Maor Farid

*Sctinium Ltd., 12 HaUmanim st., Tel-Aviv 6789731, Israel*
*Faculty of Mechanical Engineering, Technion – Israel Institute of Technology, Haifa 3200003, Israel*
*faridm@technion.ac.il*

**Abstract**

Data-Driven Response Regime Exploration and Identification (DR$^2$EI) is a novel and fully data-driven method for identifying and classifying response regimes of a dynamical system without requiring human intervention. This approach is a valuable tool for exploring and discovering response regimes in complex dynamical systems, especially when the governing equations and the number of response regimes are unknown, and the system is expensive to sample. Additionally, the method is useful for order reduction, as it can be used to identify the most dominant response regimes of a given dynamical system. DR$^2$EI utilizes unsupervised learning algorithms to transform the system's response into an embedding space that facilitates regime classification. An active sequential sampling approach based on Gaussian Process Regression (GPR) is used to efficiently sample the parameter space, quantify uncertainty, and provide optimal trade-offs between exploration and exploitation. The performance of the DR$^2$EI method was evaluated by analyzing three established dynamical systems: the mathematical pendulum, the Lorenz system, and the Duffing oscillator. The method was shown to effectively identify a variety of response regimes with both similar and distinct topological features and frequency content, demonstrating its versatility in capturing a wide range of behaviors. While it may not be possible to guarantee that all possible regimes will be identified, the method provides an automated and efficient means for exploring the parameter space of a dynamical system and identifying its underlying "sufficiently dominant" response regimes without prior knowledge of the system's equations or behavior.

*Keywords:* Response regimes identification, data-driven methods, machine learning, Gaussian process regression, uncertainty quantification.

## 1. Introduction

The study of dynamical systems is essential for understanding many natural and engineering systems, from chemical reactions to the locomotion of airspace vehicles. One of this field of research's main objectives is identifying and classifying the different response regimes of a given dynamical system [1, 2]. Discovering response regimes in a dynamical system is essential for taking precautions and preventing hazardous responses [3], such as violent structural modes. It also enables the development of reduced-order models that capture the dominant and relevant response regimes [4, 5, 6, 7, 8]. This information is crucial for designing appropriate control strategies to steer the system toward desired behaviors. However, identifying response regimes in a system typically requires either prior knowledge of its behavior or a laborious iterative process of exploring the parameters and initial conditions (ICs) space to discover potential regimes. This task can be particularly challenging and time-consuming, especially when dealing with a large number of parameters, complex systems with numerous degrees of freedom (DOFs), and unknown numbers of response regimes of a given dynamical system [9, 10, 11, 12]. In such cases, a tedious manual search of the parameter space is impractical, and grid search is often exquisitely expensive [13]. Random searches, while they may be less expensive, may overlook valuable information obtained during the search process. Furthermore, rule-based methods that rely on specific heuristics are often biased by design. For example, using a rule that targets response regimes with specific frequency characteristics, such as identifying responses with two frequencies in the fast Fourier transform (FFT), may inadvertently overlook other regimes with a broader frequency response.

Previous works have addressed the challenge of response regime identification in dynamical systems. These methods include Fourier Transform-based techniques, such as Prony's method [14], the Periodogram method [15], and the Spectral Correlation Density method [16], among others. These techniques rely on the identification of peaks in the frequency domain to infer response regimes. However, these methods are limited to systems with periodic responses, and the choice of the frequency domain resolution can significantly impact the identification process [17, 18].

Other works have proposed model-based approaches that rely on the system's governing equations to identify response regimes. These methods include bifurcation analysis, phase space analysis, and Lyapunov exponent analysis, among others [19, 2]. While these methods are effective, they require prior knowledge of the system's equations and often do not scale well to high-dimensional systems.

To address these challenges, we propose a data-driven approach that utilizes unsupervised clustering and supervised learning to identify and classify all response regimes in a dynamical system. Our method, called Data-Driven Regime Exploration and Identification (DR$^2$EI), does not require prior knowledge of the system's equations and can handle high-dimensional systems efficiently. The method is based on unsupervised learning algorithms that transform the system's response into an embedding space, enabling the classification of response regimes. The parameter space is efficiently sampled using an active sequential sampling approach [20, 21] based on Gaussian Process Regression (GPR) [22, 23, 24, 25], which enables optimal trade-offs between exploration and exploitation of the parameter and IC space. Moreover, the latter provides uncertainty quantification (UQ) regarding the existence of response regimes over the parameters space, which serves as a basis for an acquisition function of an effective active sequential search method.

In this paper, we demonstrate the effectiveness of our method using several well-known dynamical systems with different properties as benchmarks. The DR$^2$EI method successfully identified different response regimes and their corresponding separatrices and separating boundaries in the system's parameter space. The proposed method provides an automated and efficient means for exploring the parameter space of a dynamical system and identifying all of its distinct response regimes without requiring prior knowledge of the system's behavior or equations. Using those examples, we demonstrated how the DR$^2$EI method provides an automated and efficient means for exploring the parameter space of a dynamical system and identifying its most dominant response regimes without the need for prior knowledge of its behavior or human intervention. Moreover, the drawbacks and limitations of the suggested method were addressed and discussed as well.

This paper is structured as follows. Section 2 introduces the Data-Driven Regime Exploration and Identification method, including its mathematical and algorithmic formulation. In Section 3, the DR$^2$EI method is validated and applied to various widely recognized dynamical systems that possess diverse properties. The DR$^2$EI approach's results are compared with ground-truth results, which can either be analytical boundaries between the basins linked with different regimes (if accessible) or numerical results obtained in a supervised manner. Finally, in Section 4, the paper concludes by discussing the advantages and disadvantages of the DR$^2$EI method.

## 2. Data-Driven Response Regime Exploration and Identification (DR$^2$EI)

The primary goal of the suggested method is to identify all statistically significant response regimes of a dynamical system and to map their corresponding regions of the parameter space. This approach is designed to be suitable for complex systems with numerous DOFs, where access to equations of motion might be unavailable, and simulation is computationally expensive. The method is designed to be fully data-driven in order to reduce the dependence on the researcher's bias and prior knowledge or assumptions about the system's response regimes and their corresponding parameters. The objective is to develop a comprehensive and accurate understanding of the dynamical behavior of the system through a systematic and unbiased analysis of its response regimes. The DR$^2$EI method is comprised of several sequential steps.

**Step 1. Sampling the parameters space to generate an initial parameter dataset**
The first step in the DR$^2$EI method involves creating an initial dataset of the system's responses. While not mandatory, it is a good practice to do so before initiating the sequential sampling process to enrich the embedding space and facilitate the clustering process. It is recommended to allocate approximately 20% of the sampling budget to this step if the total number of samplings is limited. In this work, we choose the initial parameter dataset randomly to eliminate researcher bias toward prior knowledge and assumptions. However, other selection strategies, such as grid search, are also possible. In this study, the parameter space is denoted by $\Theta \in \mathbb{R}^n$ where $n$ is the dimension of the parameter space, and the parameter vector sampled from the parameter space is denoted by $\boldsymbol{\theta} \in \Theta \in \mathbb{R}^n$. It is essential to note that in this study, the parameter vector includes both the system's characterizing coefficients and its ICs. The sampled subset of the parameter space is denoted by $\tilde{\Theta} \in \Theta$.

**Step 2. Simulating the system and obtaining its responses corresponding to the chosen parameter sets.** Each parameter vector corresponds to a resulting response of length $n_t$, i.e. $u(t|\boldsymbol{\theta}) \in \mathbb{R}^{n_t}$. The latter is generated by running a numerical solver on the system's equations of motion (EOMs), running a computational model, or using an experimental setup.

**Step 3. Projecting the system's response to an embedding space** The system's response is projected into an embedding space $\Psi \in \mathbb{R}^m$ where $m \ll n_t$, which is a lower-dimensional representation of the high-dimensional response data. Embedding space refers to the lower-dimensional space where high-dimensional data is mapped or projected. This process retains the essential features of the original high-dimensional data, while removing any redundant or irrelevant information, making it suitable for comparisons between time histories with different time steps and lengths.

Using embedding-based approaches can be particularly useful for tasks such as clustering or classification, where the goal is to group similar data points together or assign them to specific categories. In this context, a classifier is trained on the embedding space using labeled data, and then applied to new, unlabeled data to predict its classification. By using an embedding space, the classification task becomes more computationally efficient and can potentially achieve better accuracy than working with high-dimensional data directly.

The objective of this step is to transform high-dimensional time series into a lower-dimensional embedding space while retaining the information embodied in the dynamical responses where an unsupervised clustering approach can be utilized. There are two distinct cases worth mentioning: univariate and multivariate dynamical systems.

### 2.1. Univariate dynamical system

This transformation is accomplished by utilizing a projection function or transformation, which reduces the order and maps time series from the time domain to a much lower-dimensional embedding space based on Eq. 1. Numerous techniques can be employed for reducing time series to a lower-dimensional embedding space, including FFT, which maps the time series to the frequency domain, wavelet transforms, principal component analysis (PCA), and various machine learning (ML) algorithms such as autoencoders and transformers. For clustering and classification applications, the most popular methods for time series mapping to an embedding space are Dynamic Time Warping (DTW), Singular Spectrum Analysis (SSA), Symbolic Aggregate Approximation (SAX), and Continuous Wavelet Transform (CWT). These methods allow for the representation of time series in a lower-dimensional space where they can be compared, clustered, and classified.

$$\boldsymbol{v}(\theta) = g(u(t|\boldsymbol{\theta})) \tag{1}$$

Here $g : \Theta \to \Psi$ and $\boldsymbol{v}(\boldsymbol{\theta})$ correspond to the projection function and the resulting embedding vector, respectively. In this study, we adopt Fast Fourier Transform (FFT) as the order-reduction algorithm due to its simplicity and ability to preserve the essential characteristics of the time series without requiring prior preparation or training, unlike ML-based order-reduction approaches. The FFT method maps time series from the time domain to the frequency domain, which makes it particularly useful in analyzing the oscillatory behavior of dynamical systems, as shown in Eq. 2.

$$\boldsymbol{v}(\omega|\boldsymbol{\theta}) = \int_{-\infty}^{\infty} u(t|\boldsymbol{\theta})e^{-i\omega t}dt \tag{2}$$

where $\boldsymbol{v}(\omega|\boldsymbol{\theta})$ is the Fourier transform of time series $u(t|\boldsymbol{\theta})$ in the frequency domain, and $\omega$ is the angular frequency. In this study, we utilized the discrete form of the Fourier transform, i.e. FFT, to convert the discrete-time history $u(t|\boldsymbol{\theta}) \in \mathbb{R}^{n_t}$ of the system response to an embedding vector $\boldsymbol{v}(\boldsymbol{\theta}) \in \mathbb{R}^m$, where $m$ is the dimension of the embedding vector. Since the system's response is represented by a vector, the FFT can efficiently compute the Fourier coefficients and preserve the essential features of the time series in the embedding space, as shown in Eq. 3.

$$\boldsymbol{v}(\boldsymbol{\theta}) = \text{FFT}(u[k]) = \sum_{n=0}^{N_f-1} u[n]e^{-i2\pi nk/N}, \quad k = 0, 1, \ldots, N-1 \qquad (3)$$

where $u$ is the input sequence of length $n_t$, $N_f$ is the number of frequency bins, and $k$ is the frequency index. To ensure consistency in the definition of the embedding vectors, it is important to maintain a consistent value for $N_f$.

2.2. *Multivariate dynamical system*

For multivariate dynamical systems, a response is characterized by $N$ time series of $n_t$ time steps each. The most straightforward approach is to concatenate the amplitude vector of the FFT applied on each of the DOFs' response vectors. While this approach preserves the entire frequency information about the system's response, the resulting embedding vector can be very high dimensional, on the order of $N \cdot N_f$. This approach may be suitable for small systems with few DOFs, but for larger, more complex systems it may not be computationally feasible, especially for real-time applications. Therefore, alternative dimensionality reduction methods may be more appropriate for larger systems, such as principal component analysis (PCA), can then be applied to the concatenated FFT vectors to reduce the size of the vector to the dimension of $m$, enabling effective clustering and mapping of response regimes. Alternatively, Singular Value Decomposition (SVD) can also be used to obtain the embedding vector. Another approach is to use ML methods such as autoencoders or variational autoencoders (VAEs) [26, 27] to learn a low-dimensional representation of the concatenated FFT vectors. These methods capture the nonlinear relationships between the variables and can provide more effective mapping of response regimes. Techniques such as proper orthogonal decomposition (POD) [28] or dynamic mode decomposition (DMD) [29] can also be used to reduce the dimensionality of the state space and obtain a lower-dimensional embedding vector. These methods capture the dominant modes of the system's response and allow for effective clustering and mapping of response regimes. A combined approach involves computing the FFT of each variable separately and then obtaining a low-dimensional representation of the concatenated FFT vectors using POD or DMD. This approach combines the advantages of both the FFT and the POD/DMD methods. Another approach is the multivariate singular spectrum analysis (MSSA) [30]. MSSA is a data-driven method that can decompose a multivariate time series into a set of principal components that capture the dominant spatiotemporal patterns in the data. The first few principal components can then be used as the embedding vectors representing the original time series. While MSSA captures spatiotemporal patterns in multivariate time series, it is computationally expensive and may not be suitable for large datasets or real-time applications.

**Step 4. Unsupervised regime classification by clustering in embedding space**
Once the response data has been transformed into the embedding space, the DR$^2$EI method employs unsupervised learning algorithms to identify the different response regimes of a given dynamical system. Since the time series responses are not labeled *a-priori*, and therefore the classification is based on an unlabeled dataset, unsupervised learning clustering algorithms are required. Some examples of clustering methods include k-means, hierarchical clustering, and Gaussian mixture models. In order to handle the potentially infinite number of possible response regimes, clustering methods that do not require a priori knowledge of the number of clusters are necessary. Examples of such clustering methods include density-based spatial clustering of applications with noise (DBSCAN) [31], mean-shift clustering, and spectral clustering. Among these methods, DBSCAN is a well-known and efficient approach that groups data points based on their density and has been widely used in various fields, including image analysis and computer vision.

$$C_i = \{j | d(u_i, u_j) \leq \hat{\epsilon}\} \tag{4}$$

In Eq. 4, $C_i$ - the $i$-th cluster containing all data points that are directly or indirectly reachable from point $u_i$, $j$ is an index for a data point in the dataset, $d(u_i, u_j)$ is the distance between data points $u_i$ and $u_j$, and $\hat{\epsilon}$ is a parameter defining the maximum distance between two data points for them to be considered part of the same cluster. DBSCAN algorithm involves two user-defined parameters: $\hat{\epsilon}$ and MinPts. $\hat{\epsilon}$ can be treated as the radius of the neighborhood around each point, and MinPts is the minimum number of points required to form a dense region or cluster. These parameters are important in determining the quality and accuracy of the clustering results, and their selection requires a trade-off between sensitivity and specificity. A small value of $\hat{\epsilon}$ can result in the formation of many small clusters, whereas a large value can lead to the merging of different clusters. Similarly, a small value of MinPts can result in the exclusion of some points from clusters, while a large value can cause the formation of large and dense clusters that may not be representative of the underlying data distribution. The original paper introducing DBSCAN [31] suggests that a minimum value of MinPts $= d + 1$, where $d$ is the dimensionality of the data, can be used in some cases. The selection of these parameters is typically made through a trial-and-error process or by using domain-specific knowledge.

The selection of the two parameters for the DBSCAN algorithm is one of the only interventions needed from the user. One efficient way to select these parameters is to plot a projected version of the embedding space using PCA [32] or t-SNE [33] with the DBSCAN clustering results and sample a handful of points and their corresponding time histories. If all the sampled points taken from each cluster correspond to the same response regime, both parameters are well selected. Otherwise, another trial and error iteration is needed until the user is satisfied with the clustering results.

**Step 5. Sequential Sampling and Uncertainty Quantification in Parameter Space using Gaussian Process Regression** A sequential sampling process is utilized to enable efficient exploration of the parameter space and to identify the corresponding response regimes. This is achieved by using an acquisition function that measures the expected improvement in the objective function by sampling at a particular point in the parameter space. In this study, the objective function is the behavior function denoted by $\beta(\boldsymbol{\theta})$, which is a function over the parameter space that takes integer values corresponding to different response regimes. The acquisition function balances the trade-off between exploration and exploitation and guides the selection of the next point to sample.

In this study, we use Gaussian Process Regression (GPR) to quantify the uncertainty in the parameter space regarding the behavior function. GPR is a non-parametric probabilistic regression method that provides a powerful data-driven approach to approximate unknown functions based on a set of samples. In Eq. 5 we define a latent function $\beta(\boldsymbol{\theta})$ that maps an embedding vector $\boldsymbol{v}(\boldsymbol{\theta})$ to its corresponding ground-truth value of the behavior function (the true response regime), i.e. $\beta(\boldsymbol{\theta}) : \boldsymbol{v}(\boldsymbol{\theta}) \rightarrow \tau$. Here, we assume that the latent behavior function $\beta(\boldsymbol{\theta})$ is inaccessible and can be evaluated only using noisy observations $\tau$. The noise in the output variable $\tau_i$ can be attributed to several factors, including an inadequate performance of the clustering algorithm caused by inappropriate values of its internal parameters.

Moreover, we assume that the noise is additive zero-mean and normally distributed.

$$\tau = \beta(\boldsymbol{\theta}) + \epsilon, \epsilon \sim \mathcal{N}(0, \sigma_n^2) \qquad (5)$$

Here, $\sigma_n^2$ is the variance of the noise. A Gaussian process (GP) is a collection of random variables, any finite number of which have a joint Gaussian distribution. The main idea behind GPR is to utilize a GP to represent the latent function $\beta(\boldsymbol{\theta})$. This is done in a Bayesian framework, in which the GP serves prior over function space, which in turn is conditioned on the measured data to yield a posterior distribution, as shown in Eq. 6. The latter enables us to make an inference on the values of the latent function $\beta(\boldsymbol{\theta})$ that correspond to unseen input variables using a finite set of training data $\tilde{\Theta}$.

$$\beta(\boldsymbol{\theta}) \sim \mathcal{GP}(\mu_0(\boldsymbol{\theta}), \Sigma_0(\boldsymbol{\theta}, \boldsymbol{\theta}')) \qquad (6)$$

A GP is completely specified by its mean function and covariance function, $\mu_0(\eta)$ and $\Sigma_0(\eta, \eta')$, respectively, which are given in Eq. 7

$$\mu_0(\boldsymbol{\theta}) = \mathbb{E}[\beta(\boldsymbol{\theta})], \quad \Sigma_0(\boldsymbol{\theta}, \boldsymbol{\theta}') = \mathbb{E}[\beta(\boldsymbol{\theta} - \mu_0(\boldsymbol{\theta}))\beta(\boldsymbol{\theta}' - \mu_0(\boldsymbol{\theta}'))] \qquad (7)$$

Here $\mathbb{E}$ denotes expectation. The mean functions $\mu_0(\boldsymbol{\theta})$ of the prior GP embodied a prior knowledge about the latent function $\beta(\boldsymbol{\theta})$. Usually, it is unknown and therefore set to zero. The properties of the GP are governed by the covariance function or kernel, which is symmetric and positive semi-definite by definition. The covariance function $\Sigma_0(\boldsymbol{\theta}, \boldsymbol{\theta}')$ quantifies the covariance between any pair of points in the dataset with respect to a given covariance function. Among the most commonly used kernels is the squared exponential kernel, also known as the Radial Basis Function (RBF) kernel or Gaussian kernel, which is defined as follows:

$$\Sigma_0(\boldsymbol{\theta}, \boldsymbol{\theta}') = \sigma_l^2 \exp\left(-\frac{(\boldsymbol{\theta} - \boldsymbol{\theta}')^2}{2l^2}\right) \qquad (8)$$

Here parameter $l$ is a characteristic length scale that repents the covariance between a data point to its neighbors and thus dictates the smoothness of the candidate functions generated by the GP. For example, large values of $l$ enforce large off-diagonal values in the covariance matrix, which corresponds to smooth candidate function, and vice versa. In general, extrapolation using GPR is reliable, only approximately $l$ units away from the training data. Parameter $\sigma_l$ determines the variance of the candidate functions away from the mean $\mu_0(\boldsymbol{\theta})$. The set of parameters $\boldsymbol{\gamma} = \{\sigma_n, l, \sigma_l\}$, also called hyper-parameters, are also determined by the user to best fit the data by optimizing over log-likelihood loss function according to the maximum *a-posteriori* (MAP) principle, as shown in Eq. 9:

$$\mathcal{L}_{\text{MAP}}(\boldsymbol{\theta}) = log(\tau|\boldsymbol{\gamma}) = \frac{1}{2}\boldsymbol{\tau}^T(K_{\boldsymbol{\tau},\boldsymbol{\tau}})^{-1}\boldsymbol{\tau} - \frac{1}{2}\log K_{\boldsymbol{\tau},\boldsymbol{\tau}} - \frac{1}{2}N_{\text{CV}}\log 2\pi \qquad (9)$$

Here, $K$ is the symmetric covariance matrix whose $ij^{th}$ entry is the covariance between the $i^{th}$ variable in the group denoted by the first subscript and the $j^{th}$ variable in the group denoted by the second subscript, calculated using covariance function $\Sigma_0$ and corresponding hyper-parameters. Vector $\boldsymbol{\tau}$ is a vector of the training observations, and $K_{\boldsymbol{\tau},\boldsymbol{\tau}} \equiv K_{\mathbf{g},\mathbf{g}} + \sigma_n \mathbb{I}$.

The vector of optimal hyper-parameters $\hat{\boldsymbol{\gamma}}$ is obtained by optimization of the log-likelihood loss function over the hyperparameter space $\boldsymbol{\Gamma}$, in accordance to Eq. 10:

$$\hat{\boldsymbol{\gamma}} = \arg\min_{\boldsymbol{\gamma} \in \boldsymbol{\Gamma}} \mathcal{L}_{\text{MAP}}(\boldsymbol{\gamma}) \tag{10}$$

After the optimal hyper-parameters $\hat{\boldsymbol{\gamma}}$ over the training data are obtained, the GPR is fully-determined. The inference can be easily made by computing the posterior multivariate PDF obtained after conditioning on the training data, which in our case is the sampled parameter sets $\tilde{\Theta}$ and their corresponding behavior function values $\beta(\boldsymbol{\theta})$, i.e., $\mathcal{D} = \{\boldsymbol{\theta}_i, \tau_i\}, \boldsymbol{\theta}_i \in \tilde{\Theta}$, as shown in Eq. 11

$$\boldsymbol{\tau}^* | \mathcal{D}, \boldsymbol{\theta}^* \sim \mathcal{N}(\boldsymbol{\mu}(\boldsymbol{\theta}^* | \mathcal{D}), \boldsymbol{\Sigma}(\boldsymbol{\theta}^* | \mathcal{D})) \tag{11}$$

Here, $\boldsymbol{\tau}$ is the estimated observations vector that corresponds to the vector of test inputs $\beta(\boldsymbol{\theta})$, for which we want to estimate the latent function $\beta(\boldsymbol{\theta})$. $\boldsymbol{\mu}(\boldsymbol{\theta}^* | \mathcal{D})$ and $\boldsymbol{\Sigma}(\boldsymbol{\theta}^* | \mathcal{D}))$ are the mean and covariance function of the posterior GP, obtained after conditioning the prior of the training data $\mathcal{D}$, as given in Eq. 12:

$$\begin{aligned}\boldsymbol{\mu}(\boldsymbol{\theta}^* | \mathcal{D}) &\equiv \mathbb{E}\left[\boldsymbol{\tau}^* | \mathcal{D}, \boldsymbol{\theta}^*\right] = K(\boldsymbol{\tau}^*, \boldsymbol{\tau}) \left(K(\boldsymbol{\tau}, \boldsymbol{\tau}) + \sigma_n^2 \mathbb{I}\right)^{-1} \boldsymbol{\tau} \\ \boldsymbol{\Sigma}(\boldsymbol{\theta}^* | \mathcal{D}) &= K(\boldsymbol{\theta}^*, \boldsymbol{\theta}^*) - K(\boldsymbol{\theta}^*, \boldsymbol{\theta}) \left(K(\boldsymbol{\theta}, \boldsymbol{\theta}) + \sigma_n^2 \mathbb{I}\right)^{-1} K(\boldsymbol{\theta}, \boldsymbol{\theta}^*)\end{aligned} \tag{12}$$

The inherent UQ provided by GPR, represented by $\boldsymbol{\Sigma}(\boldsymbol{\theta}^* | \mathcal{D})$, enables us to prioritize sampling at points where the uncertainty in the value of the behavior function $\beta(\boldsymbol{\theta})$ (and thus, in the response regime classification in a given volume in the parameter space) is highest. The goal of the DR$^2$EI method is to identify iteratively and sample points in the parameter space that are most informative, leading to a more efficient exploration of the parameter space and a more accurate identification of all response regimes of the system.

To implement the acquisition function, we compute the expected improvement (EI) at each point in the parameter space. The EI acquisition function is defined as the improvement in the behavior function that we expect to obtain by sampling at a particular point $\boldsymbol{\theta}$ in the parameter space relative to the current best estimate of the maximum behavior function value. In other words, it measures how much better the behavior function will be defined is expected to be at a given point compared to the current best response regime identified.

The EI is calculated as follows:

$$\begin{aligned} EI(\boldsymbol{\theta}|\mathcal{D}) &= \begin{cases} (\tau(\boldsymbol{\theta}|\mathcal{D}) - \beta_{max} - \zeta)\Phi(Z) + \sigma(\boldsymbol{\theta}|\mathcal{D})\phi(Z), & \text{if } \sigma(\boldsymbol{\theta}) > 0 \\ 0, & \text{otherwise} \end{cases} \\ \tau(\boldsymbol{\theta}) &= \boldsymbol{\mu}(\boldsymbol{\theta}|\mathcal{D}) \\ \boldsymbol{\sigma}(\boldsymbol{\theta}|\mathcal{D}) &= \sqrt{\text{diag}\left(\boldsymbol{\Sigma}(\boldsymbol{\theta}|\mathcal{D})\right)} \end{aligned} \tag{13}$$

Where diag is an operator that returns the diagonal of a matrix, $\tau(\boldsymbol{\theta})$ is the behavior function value at point $\boldsymbol{\theta}$ obtained from the GPR, $\beta_{max}$ is the current best estimate of the maximum objective function value, $\sigma(\boldsymbol{\theta})$ is the standard deviation of the GPR at point $\boldsymbol{\theta}$, $\Phi(Z)$ and $\phi(Z)$ are the cumulative distribution function (CDF) and probability density function (PDF) of the standard normal distribution, respectively, and $Z$ is a standard normal random variable. The parameter $\zeta$ controls the trade-off between exploration and exploitation, with higher values favoring exploration and lower values favoring exploitation, as given in Eq.14:

$$\boldsymbol{\sigma}(\boldsymbol{\theta}|\mathcal{D}) \leq \zeta \qquad (14)$$

The intuition behind the EI function is that it favors sampling at points with both high behavior function values and high uncertainty, as statistically, these points have the potential to improve the current best estimate of the most exotic response regimes. This is due to the stochastic nature of the initial random sampling of the parameter space, which means that the less probable response regimes may not be identified in the initial random sampling or may only be found after the more common regimes that correspond to larger volumes in the parameter space. As DBSCAN allocates increasing integers to new clusters it encounters, the less probable regimes will likely correspond to higher integers.

**Step 6. Iterative Refinement of Inter-Regime Separation Boundaries Using Gaussian Process Regression** As part of the iterative process of sequentially sampling the parameter space to identify additional response regimes, the separation boundaries between hypervolumes in the parameter space corresponding to distinct response regimes are iteratively refined. As mentioned earlier, the mean function of the GPR, denoted as $\boldsymbol{\mu}(\boldsymbol{\theta}|\mathcal{D})$, serves as an approximation to the latent behavior function $\beta(\boldsymbol{\theta})$, and this approximation improves with each sampling iteration. The separation boundaries in the parameter space, which distinguish between hypervolumes that correspond to different response regimes, can be expressed as the following series: $\boldsymbol{\mu}(\boldsymbol{\theta}|\mathcal{D}) = \frac{1}{2}, 1\frac{1}{2}, ..., N_r - \frac{3}{2}$, where $N_r$ denotes the number of distinct response regimes (or clusters) identified by the clustering algorithm, such as DBSCAN in the current study.

Steps $2 - 6$ involve selecting the parameter set with the highest EI value, simulating the system for this parameter set, projecting the resulting response to the embedding space, clustering the resulting points with DBSCAN, classifying the corresponding response regime, and refining the inter-regime separation boundaries. By repeating those steps iteratively, we can efficiently explore the parameter space to identify the system's various response regimes, while the accuracy of the partition boundaries between hypervolumes corresponding to distinct response regimes is improved with each iteration.

Ideally, we will continue sampling the parameter space until the maximum value of the standard deviation of the GPR falls below the desired uncertainty threshold $\zeta$. This threshold defines the level of certainty in the exploration of the parameter space, which allows for mapping the parameter space to different regimes volumes and discovering all possible response regimes. Alternatively, if a maximum number of iterations $T_{max}$ is given, the stopping criterion will be to run $T_{max}$ iterations of the iterative process. Thus, the set of design parameters that define the sensitivity of the DR$^2$EI method and must be selected by the user a priori is denoted as follows: $\Delta = \{\sigma_n, l, \sigma_l, \hat{\epsilon}, MinPts, \zeta\}$. Additionally, the user must also choose the region of investigation in the parameter space, denoted as $\partial\tilde{\Theta}$. Algorithm 1 presents the DR$^2$EI Method with an Uncertainty-Based Stopping Criterion, while Algorithm 2 presents the DR$^2$EI Method with a Maximum Iteration Limit.

**Input:** Vector of user-defined sensitivity parameters $\Delta$, initial random sampling set $\Theta_0$ within a chosen domain in the parameter space $\partial \tilde{\theta}$, acquisition function $A$, Gaussian Process model $M$, uncertainty threshold $\zeta$

**Output:** Estimated behavior function $\beta(\boldsymbol{\theta})$

$\mathcal{D} \leftarrow \{(\boldsymbol{\theta}, \tau(\boldsymbol{\theta})) \mid \boldsymbol{\theta} \in \Theta_0\}$
$M.\text{fit}(\mathcal{D})$;
$\tilde{\Theta} \leftarrow \Theta_0$;
**while** *True* **do**
$\quad$ $\boldsymbol{\theta}_{t+1} \leftarrow \arg\max_{\boldsymbol{\theta} \notin \tilde{\Theta}} A(\boldsymbol{\theta}|M)$;
$\quad$ $\tau_{t+1} \leftarrow \text{evaluate}(\boldsymbol{\theta}_{t+1})$;
$\quad$ $\mathcal{D} \leftarrow \mathcal{D} \cup (\boldsymbol{\theta}_{t+1}, \tau_{t+1})$;
$\quad$ $M.\text{fit}(\mathcal{D})$;
$\quad$ $\tilde{\Theta} \leftarrow \tilde{\Theta} \cup \boldsymbol{\theta}_{t+1}$;
$\quad$ $\beta(\boldsymbol{\theta}) \leftarrow M.\text{predict}(\boldsymbol{\theta})$;
$\quad$ **if** $\max \sigma(\boldsymbol{\theta}) < \zeta, \boldsymbol{\theta} \in \Theta$ **then**
$\quad\quad$ | break;
$\quad$ **end**
**end**

**Algorithm 1:** The DR$^2$EI method with uncertainty-based stopping criterion

**Input:** Vector of user-defined sensitivity parameters $\Delta$, Initial random sampling set $\Theta_0$ within a chosen domain in the parameter space $\partial \tilde{\theta}$, maximum number of iterations $T\max$, acquisition function $A$, Gaussian Process model $M$

**Output:** Estimated behavior function $\beta(\boldsymbol{\theta})$

$\mathcal{D} \leftarrow \{(\boldsymbol{\theta}, \tau(\boldsymbol{\theta})) \mid \boldsymbol{\theta} \in \Theta_0\}$
$M.\text{fit}(\mathcal{D})$;
$\tilde{\Theta} \leftarrow \Theta_0$;
**for** $t \leftarrow 1$ **to** $T_{\max}$ **do**
$\quad$ $\boldsymbol{\theta}_{t+1} \leftarrow \arg\max_{\boldsymbol{\theta} \notin \tilde{\Theta}} A(\boldsymbol{\theta}|M)$;
$\quad$ $\tau_{t+1} \leftarrow \text{evaluate}(\boldsymbol{\theta}_{t+1})$;
$\quad$ $\mathcal{D} \leftarrow \mathcal{D} \cup (\boldsymbol{\theta}_{t+1}, \tau_{t+1})$;
$\quad$ $M.\text{fit}(\mathcal{D})$;
$\quad$ $\tilde{\Theta} \leftarrow \tilde{\Theta} \cup \boldsymbol{\theta}_{t+1}$;
$\quad$ $\beta(\boldsymbol{\theta}) \leftarrow M.\text{predict}(\boldsymbol{\theta})$;
$\quad$ **if** $\max \sigma(\boldsymbol{\theta}) < \zeta, \boldsymbol{\theta} \in \Theta$ **then**
$\quad\quad$ | break;
$\quad$ **end**
**end**

**Algorithm 2:** The DR$^2$EI method with maximum iteration Limit $T_{max}$

## 3. Numerical Validation

The goal of this section is to evaluate the effectiveness of the DR$^2$EI method in identifying the various dynamical regimes of a given system and generating accurate inter-regime separation boundaries, without the need for prior knowledge. Therefore, we conducted a comprehensive benchmarking study on three well-known and fundamental dynamical systems. These systems possess distinct characteristics and underlying dynamical regimes. The first system considered is the mathematical pendulum, which is a univariate second-order dynamical system. It exhibits two distinct dynamical regimes in state space, characterized by oscillations and rotations with notably different topological properties. The separatrix between these regimes is given analytically. The second system is the Lorenz system, which is a multivariate three-dimensional and chaotic dynamical system. It exhibits diverse behaviors such as steady-state, periodic, and chaotic responses. Finally, we examined the Duffing oscillator, which is a univariate second-order system displaying two dynamical regimes that possess identical topological characteristics, namely small and high amplitude responses.

3.1. Univariate dynamical system- the mathematical pendulum

In this benchmark, we took the ICs of the pendulum as the only parameters for the sake of simplicity and visuality of the demonstration, i.e., $\boldsymbol{\theta} = \{x_0, v_0\} \in \mathbb{R}^2$ and take all other systems parameters as fixed $\bar{\omega} = 1, \bar{\lambda} = 0$. We apply DR$^2$EI's sequential search algorithm for sets of ICs between the ranges $x_0 \in [-3.5, 3.5], \dot{\theta} \in [-2.5, 2.5]$, and simulate the corresponding system's responses. Each time history is simulated using the system's EOM as shown in Eq. 15 for a termination time of $t_f = 200$ and $n_t = 1000$ time steps.

$$\ddot{x} + \bar{\lambda}\dot{x} + \bar{\omega}^2 x = 0, \ \ x(0) = x_0, \dot{x}(0) = v_0 \qquad (15)$$

The data obtained from the simulations were then transformed into an $m = N_f$-dimensional embedding space using FFT. In this analysis, we take $N_f = 1024$. To visualize the transformed vectors in the embedding space, we employed the PCA method to project the embedding vectors onto a 2D space, as shown in Fig.1. As mentioned above, PCA is a statistical method that can identify patterns in data and summarize the information content of a large number of variables into a smaller number of principal components. In this context, PCA was used to identify the most significant directions in the embedding space and project the high-dimensional vectors onto a 2D plane while preserving the relative distances between them. Close points in the embedding space were transformed to close points in the 2D plane, and vice versa for far points. Using the DBSCAN algorithm, we were able to cluster these regimes in an unsupervised fashion and classify them accordingly.

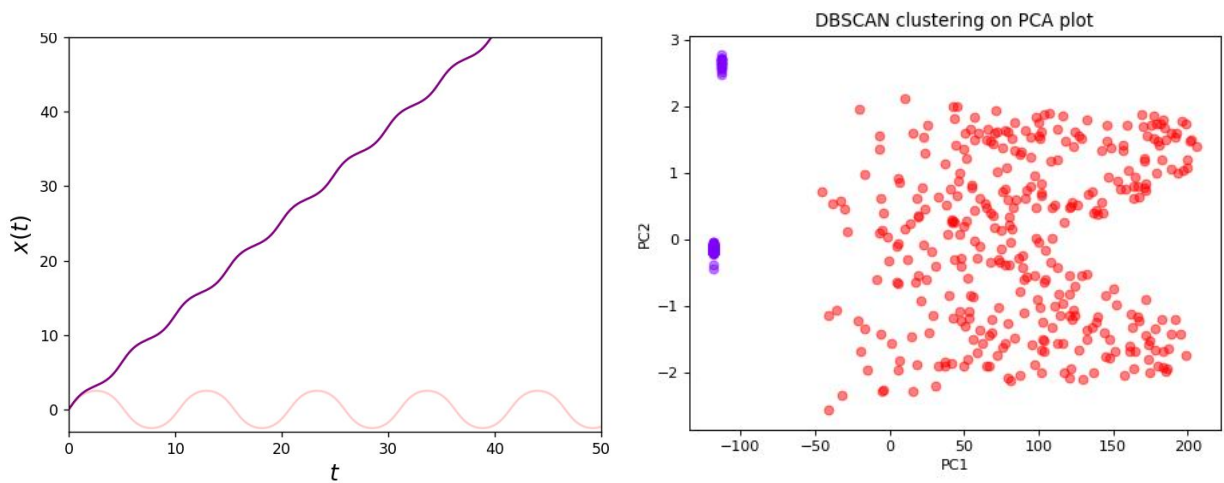

Figure 1: Left- Illustration of the response regimes of the mathematical pendulum, which were identified through parameter space exploration using the DR$^2$EI method, including oscillatory response (red) and rotational response (purple). Right- Projection of the embedding space onto a two-dimensional plane using Principal Component Analysis (PCA). The dots with different colors correspond to time series that were clustered using DBSCAN into different response regimes; for parameters: $\hat{\epsilon} = 7$ and MinPts $= 5$.

Fig. 1 shows the clustering of all system responses into two distinct regimes: oscillatory (red dots) and rotational (purple dots). However, the rotational regime appears to be divided into two separate sub-clusters upon closer examination. Further analysis reveals that these sub-clusters correspond to rotations in opposite directions (left and right).

We incorporate the results obtained from the DBSCAN clustering algorithm into the behavior function $\beta(\boldsymbol{\theta})$ defined over the parameter space. Subsequently, we train a GPR model on the resulting behavior function. The mean function of the trained GPR captures the underlying latent behavior function of the system, while the standard deviation represents the uncertainty in the behavior function's value for a given parameter set. To further explore the parameter space, we apply sequential sampling within the predefined boundaries. At each iteration, we sample the parameter space $\Theta$ using the parameter set $\boldsymbol{\theta}$ that corresponds to the highest EI value.

We assess the accuracy of the DR$^2$EI method in identifying response regimes by plotting the classified points in their corresponding locations on the parameter (ICs) plane and comparing them to the analytical separatrix given in Eq. 16, which separates the two dynamical regimes of rotations and oscillations. The results show that DR$^2$EI accurately identified and classified all response regimes in the parameter space without a significant human intervention, except for defining the algorithm's sensitivity parameters $\Delta$.

$$v_0 = \pm\sqrt{2(1 - cos(x_0))} \qquad (16)$$

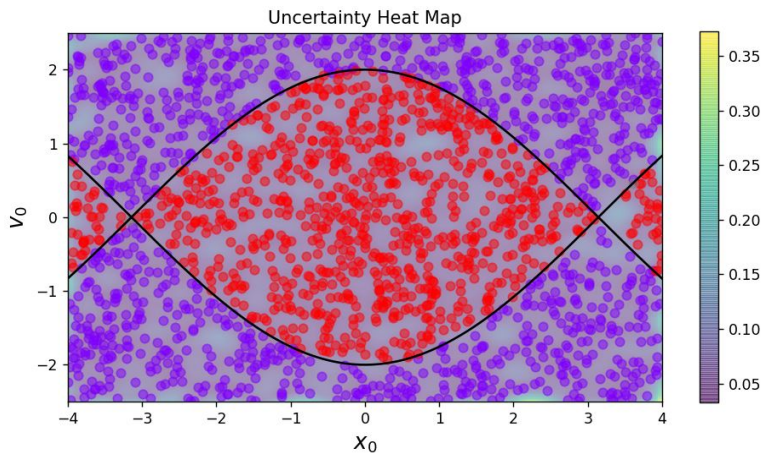

Figure 2: The parameter sets sampled for the mathematical pendulum with parameters $\bar{\omega} = 1$ and $\bar{\lambda} = 0$, colored by their class according to the DR$^2$EI method. The heat-map represents the GPR standard deviation, which serves as an UQ measure for the classification of response regimes over the parameter space. The oscillatory and rotational regimes are indicated by red and purple dots, respectively. Furthermore, the black line represents the analytical expression of the separatrix that corresponds to Eq.16.

The resulting heat-map shown in Fig. 3 represents the uncertainty measure over the parameter space, which was minimized to guide the sampling process.

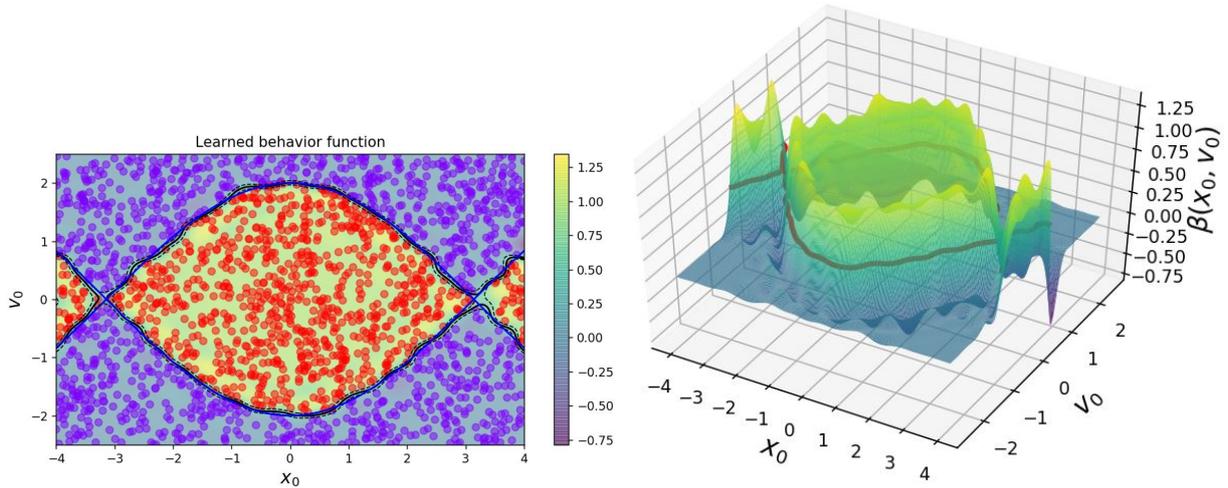

Figure 3: Left- Performance evaluation of DR$^2$EI method over the parameter (ICs) space of the mathematical pendulum with parameters $\bar{\omega} = 1$ and $\bar{\lambda} = 0$. The learned behavior function, which corresponds to the mean function of the GPR, is plotted over the IC plane. Purple and red dots indicate the sample parameters corresponding to the rotational and oscillatory regimes, respectively. Both regimes were classified based on the learned behavior function values of zero and one, respectively. The blue line represents the analytical expression of the separatrix. The solid black line in the figure represents the estimated boundary of the response regimes identified by the DR$^2$EI method, which corresponds to the contour $\beta(\boldsymbol{\theta}) = 0.5$. The inner and outer black dashed lines indicate the boundaries of the response regimes corresponding to $\beta(\boldsymbol{\theta}) = 0.325$ and $\beta(\boldsymbol{\theta}) = 0.675$, respectively. Right- 3D view of the learned behavior function $\beta(\boldsymbol{\theta})$ over the IC plane, and the separation contour (solid red line) that corresponds to $\beta(\boldsymbol{\theta}) = 0.5$.

## 3.2. Multivariate dynamical system- The Lorenz system

In this section, we apply the DR$^2$EI method to the Lorenz System, a well-known multivariate three-dimensional dynamical system that exhibits chaotic behavior for some parameters, as described in Eq. 3.2. Since this system has only a few degrees of freedom and we aim to preserve as much information as possible about the system's dynamical response, we employ the straightforward approach of concatenating the amplitude vectors of the FFT applied to each time history of the three degrees of freedom of the system. For visualization and pedagogical reasons, we fix two of the system's parameters and three of its initial conditions and study the possible response regimes that exist on the parameter plane of $x_0$ and $\rho$.

$$\begin{cases} \dfrac{dx}{dt} = s(y - x) \\ \dfrac{dy}{dt} = x(\rho - z) - y \\ \dfrac{dz}{dt} = xy - bz \end{cases} \qquad (17)$$

where $x$, $y$, and $z$ are the state variables, and $s$, $\rho$, and $b$ are the system parameters. Varaible $x$ represents the rate of convective overturning, $y$ represents the horizontal temperature variation, and $z$ represents the vertical temperature variation. Parameter $b$ is known as the Prandtl number. Parameter $s$ represents the rate of heat transfer and is related to the ratio of the fluid velocity and the temperature difference. The parameter $\rho$ represents the ratio of the width to the height of the convecting layer and controls the behavior of the system. When $\rho$ is greater than a certain critical value, the system exhibits chaotic behavior.

We apply the DR$^2$EI method to explore, map, and identify the response regimes that "live" in the $x_0 - \rho$ plane of the Lorenz system. Specifically, we consider the parameter values $s = 10$ and $b = 8/3$, and initial conditions of $y_0 = 0, z_0 = 0$. We concatenate the time series FFTs into an $m = N \cdot N_f$ dimensional embedding vector. In Fig. 4, we present the projection of the resulting embedding space and observe a clear distinction between two response regimes: the purple corresponds to steady-state response, i.e., the system converges to a fixed value, while the remaining points correspond to chaotic responses.

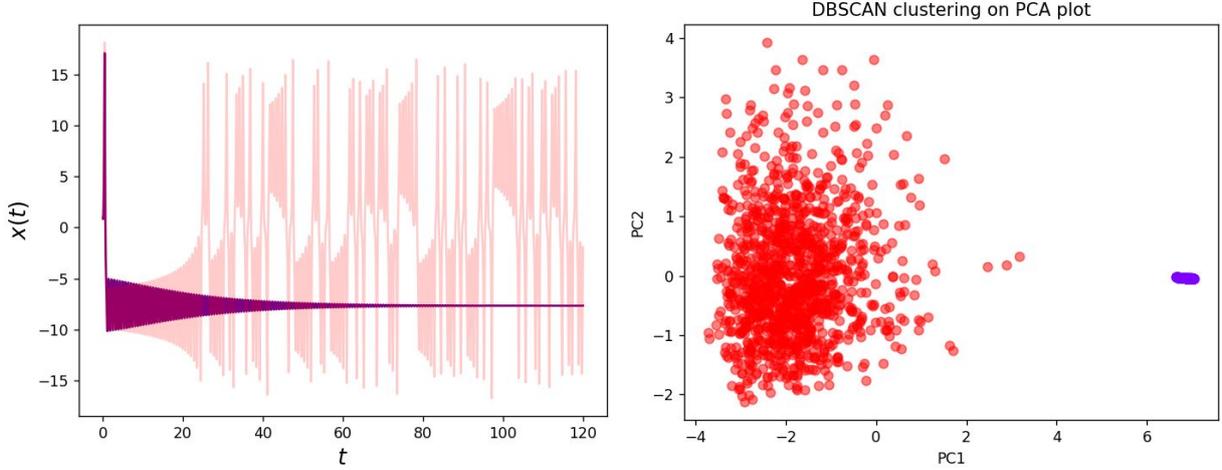

Figure 4: Left- Illustration of the response regimes of the Lorenz system, which were identified through parameter space exploration using the DR$^2$EI method, including steady-state response (purple) and chaotic response, converging to a strange attractor in state space (red), for parameters $s = 10, b = 8/3, x_0 = 1, y_0 = 0, z_0 = 0$ and $\rho = 23$ (steady-state response) and $\rho = 25$ (chaotic response). Right- Projection of the embedding space onto a two-dimensional plane using Principal Component Analysis (PCA). The dots with different colors correspond to time series that were clustered using DBSCAN into different response regimes; for parameters: $\hat{\epsilon} = 5$ and MinPts = 4.

In Fig. 5, the classified sampled parameter sets are plotted over the parameters plane. The heat map corresponds to the measure of the UQ in the underlying response regime, i.e., the variance of the GPR. The homogeneity of the UQ level means that sequential sampling was satisfactory and reduces the uncertainty regarding the existing response regimes evenly across the parameter plane. The black contour corresponds to $\beta(\boldsymbol{\theta}) = 0.5$, which is the boundary between the areas corresponding to a zero value, corresponding to steady-state responses, and one value, corresponding to chaotic behavior.

It is worth noting that even though the system converges to different fixed values depending on the parameters, the DR$^2$EI method still classifies them as the same response regime, as intended.

In Fig. 6, we present the distribution of the behavior function over the parameter plane, where zero and one correspond to steady state response and chaotic response, respectively. Like in Fig. 5, the black contour corresponds to $\beta(\boldsymbol{\theta}) = 0.5$. As one can see, the method successfully mapped the response regimes and fits the boundary between their corresponding regions in the parameter space.

However, Fig. 6 highlights a limitation of the DR$^2$EI method: it can only identify domains with a notable or significantly large hyper-volume in the parameter space. A "notable" volume is typically defined as one that is sampled at least MinPts times in random sampling to ensure that the regimes are clustered in the embedding space. For instance, in this figure, no chaotic response is observed for any $\rho$ value and $x_0 = 0$. However, since this is a line with no width by definition, it does not possess any volume in the parameter space, and therefore, it is not captured by the DR$^2$EI method. Consequently, it does not have any impact on the behavior function.

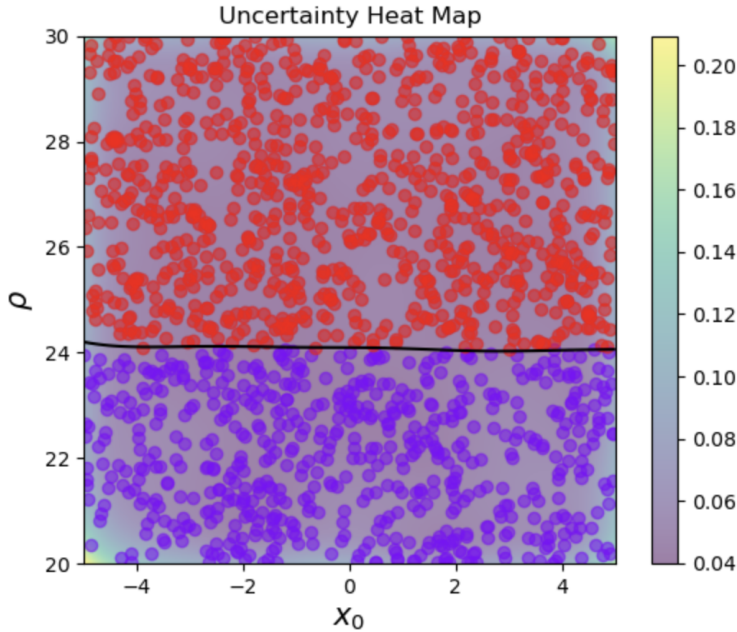

Figure 5: The parameter sets sampled for the Lorenz system with parameters $s = 10$ and $b = 8/3$, and initial conditions of $y_0 = 0, z_0 = 0$, colored by their class according to the DR$^2$EI method. The heat-map represents the GPR standard deviation, which serves as an UQ measure for the classification of response regimes over the parameter space. The parameter sets that correspond to steady-state and chaotic regimes are indicated by red and purple dots, respectively. Furthermore, the black line represents DR$^2$EI's estimated boundary between distinct regimes, i.e. $\beta(\boldsymbol{\theta}) = 0.5$.

### 3.3. The Duffing oscillator- multiple periodic solutions

The Duffing oscillator is a second-order dynamical system with a single degree of freedom that exhibits multiple solutions due to its cubic nonlinearity, as shown in Eq. 18.

$$\ddot{x} + 2\bar{\epsilon}\bar{\lambda}\dot{x} + x + \bar{\epsilon}\bar{\alpha}x^3 = \bar{\epsilon}\bar{F}\sin\left((1 + \bar{\epsilon}\bar{\sigma})t\right) \tag{18}$$

In this equation, $x$ represents the displacement of the oscillator, while $\bar{\epsilon}$, $\bar{\sigma}$, $\bar{F}$, $\bar{\lambda}$, and $\bar{\alpha}$ denote the small parameter, detuning parameter, external excitation amplitude, damping coefficient, and cubic nonlinearity coefficient, respectively. The coexistence of solutions corresponds to the bending of the frequency response curve, resulting in either hardening or softening behavior. The frequency response curve of the Duffing equation, as given by Equation 19, yields the steady-state amplitudes of its two stable solutions. Figure 7 shows both the frequency response curve and the periodic responses that correspond to the same parameter set. In this context, the dynamical regimes that arise in response to the oscillator can be classified based on the amplitude of their responses. These distinct regimes exhibit identical topological structures and frequency content but differ only in their response amplitudes, which correspond to differences in the amplitude of their frequency spectra.

$$\bar{\sigma} = \frac{3\bar{\alpha}}{8}a^2 \pm \sqrt{\frac{\bar{F}^2}{4a^2} - \bar{\lambda}^2} \tag{19}$$

Here, $a$ represents the steady-state amplitude of the oscillator's response. In contrast to the Duffing oscillator, the dynamical regimes observed in the previous examples were different not only in their frequency content and time response but also in the topology of their underlying attractors. These attractors included points, contours, or strange attractors. To better understand the distinct regimes exhibited by the Duffing oscillator, we employed the DR$^2$EI algorithm, which mapped the corresponding basins in parameter space. The results of this analysis are presented in Fig. 8-10.

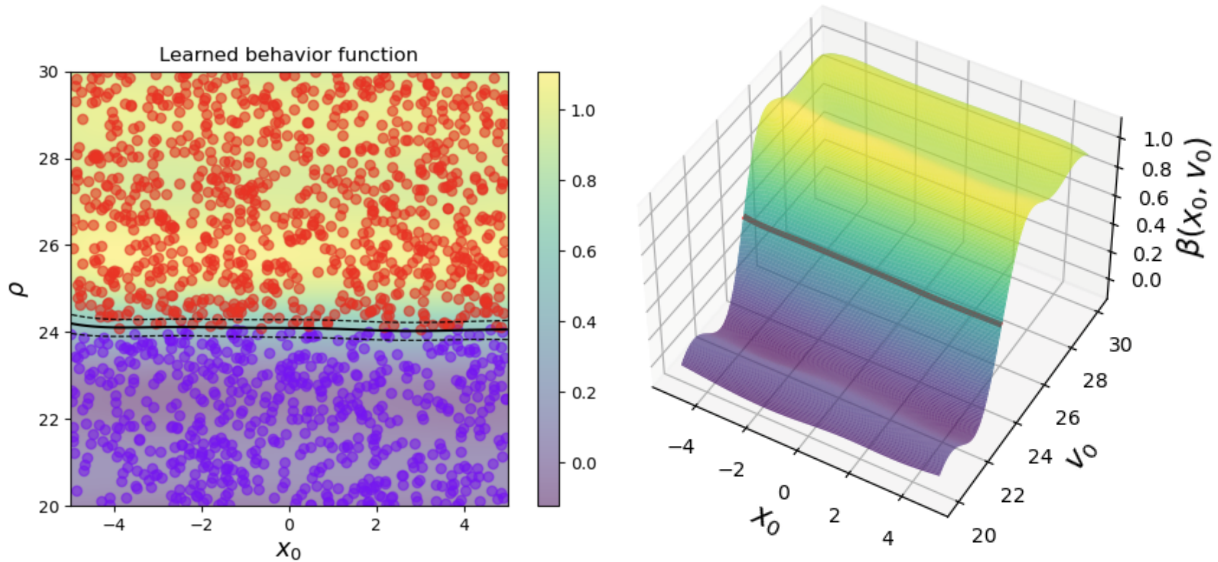

Figure 6: Left- Performance evaluation of DR$^2$EI method over the parameter (ICs) space of the Lorenz system with parameters $s = 10$ and $b = 8/3$, and initial conditions of $y_0 = 0, z_0 = 0$. The learned behavior function, which corresponds to the mean function of the GPR, is used to identify the classified sampled points from the parameter space. Purple and red dots indicate the sample parameters corresponding to the chaotic and steady-state regimes, respectively. Both response regimes were classified based on the behavior function values of zero and one, respectively. The solid black line represents the estimated boundary of the response regimes identified by the DR$^2$EI method, which corresponds to $\beta(\boldsymbol{\theta}) = 0.5$. The lower and upper black dashed lines indicate the boundaries of the response regimes corresponding to $\beta(\boldsymbol{\theta}) = 0.325$ and $\beta(\boldsymbol{\theta}) = 0.675$, respectively. Right- 3D view of the learned behavior function $\beta(\boldsymbol{\theta})$, and the separation contour (red solid line) that corresponds to $\beta(\boldsymbol{\theta}) = 0.5$.

## 4. Concluding remarks

In this paper, we presented a novel data-driven method for identifying and classifying response regimes in dynamical systems using unsupervised clustering and supervised learning. The method is fully automated, requires no prior knowledge of the system's behavior or governing equations, and allows for the exploration of the parameter space of complex systems where traditional methods are infeasible.

The proposed method is particularly useful for systems where manual exploration of the parameter space is a tedious and resource-intensive task. The method allows for efficient and automated exploration of the parameter space, resulting in the identification of "statistically-significant" response regimes in a fully data-driven fashion. The method uses unsupervised learning algorithms to transform the system's response into an embedding space, where response regimes can be classified using supervised learning algorithms. The method's active sequential sampling approach based on Gaussian Process Regression allows for optimal trade-offs between exploration and exploitation, enabling efficient sampling of the parameter space. By utilizing the mean function of the Gaussian Process Regression (GPR), one can learn the separation boundaries in the parameter space that distinguish hypervolumes corresponding to different response regimes.

The suggested method was implemented on several well-known dynamical systems, including the mathematical pendulum (univariate), Lorenz system (multivariate), and the Duffing os-cillator (coexistence of multiple periodic solutions with similar topology).

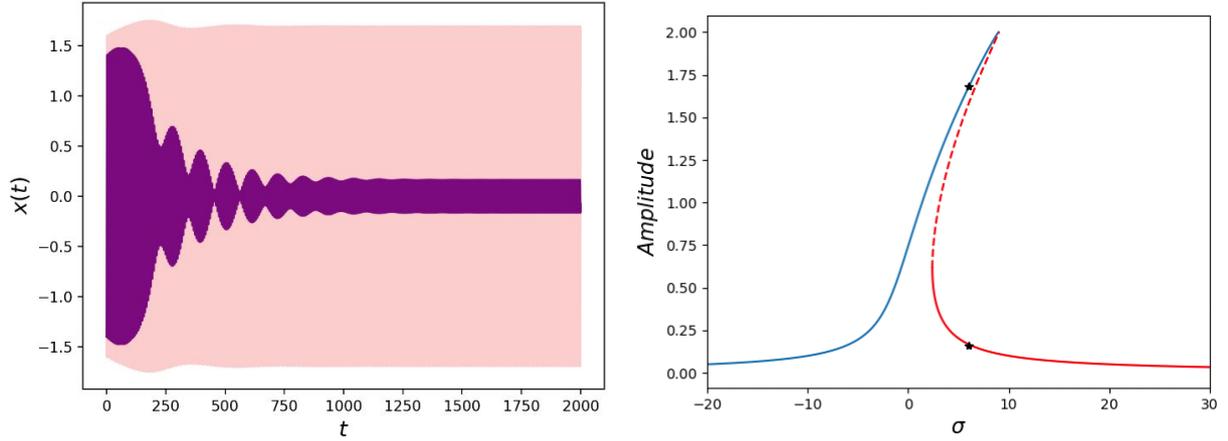

Figure 7: Left- Illustration of the response regimes of the Duffing oscillator, which were identified through parameter space exploration using the DR$^2$EI method, including small-amplitude response (purple) and high-amplitude response (red) for parameters: $\epsilon = 0.01, \bar{\sigma} = 6, \bar{F} = 2, \bar{\alpha} = 6, \bar{\lambda} = 0.5, \dot{x}_0 = 0$ and $x_0 = -1.4$ (purple) $x_0 = -1.6$ (red). Right- The underlying frequency response curve; solid lines and dashed lines correspond to stable and unstable solutions, respectively. The red and blue branches correspond to the plus and minus signs in Eq. 19, respectively. Black stars correspond to the possible steady-state responses for $\bar{\sigma} = 6$.

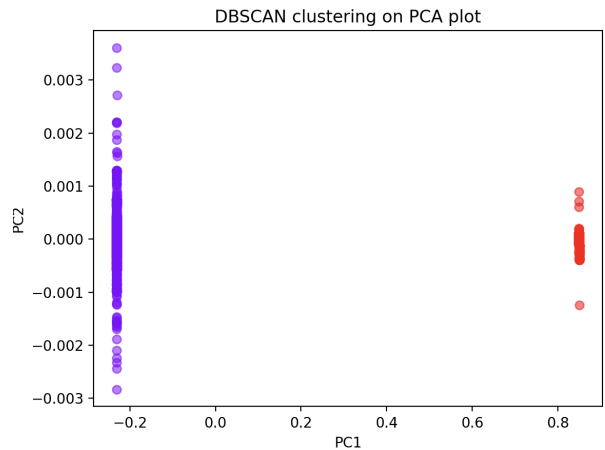

Figure 8: Projection of the embedding space onto a two-dimensional plane using Principal Component Analysis (PCA). The dots with different colors correspond to time series that were clustered using DBSCAN into different response regimes; for parameters: $\hat{\epsilon} = 5$ and MinPts $= 4$.

Our results showed that DR$^2$EI was able to accurately identify and classify different response regimes in all examples. Notably, in the case of the pendulum system, where the separatrix is given analytically, we compared the identified regimes to the analytical separatrix and found good agreement. These findings demonstrate the effectiveness of DR$^2$EI in identifying distinct response regimes, even in the absence of analytical knowledge of the system's behavior.

The proposed method has several advantages over traditional methods. Firstly, it requires no prior knowledge of the system's behavior or governing equations, making it suitable for complex systems where obtaining such information is difficult or impossible. Secondly, it allows for efficient and automated exploration of the parameter space, resulting in the identification of all response regimes in a fully data-driven fashion. Finally, the method allows for the efficient sampling of the parameter space using an active sequential sampling approach based on Gaussian Process Regression, enabling optimal trade-offs between exploration and exploitation.

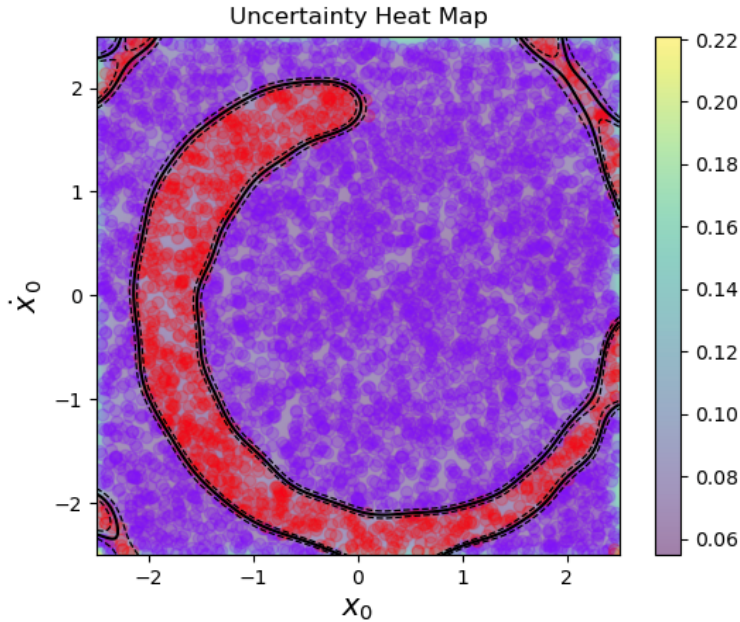

Figure 9: the parameter sets sampled for the Duffing oscillator with parameters $\bar{\epsilon} = 0.01, F = 2, \bar{\alpha} = 6$ and $\bar{\lambda} = 0.5$ colored by their class according to the DR$^2$EI method. The heat-map represents the GPR standard deviation, which serves as an UQ measure for the classification of response regimes over the parameter space. The parameter sets that correspond to high- and low-amplitude periodic solutions are indicated by red and purple dots, respectively. Furthermore, the black line represents DR$^2$EI's estimated boundary between distinct regimes, i.e. $\beta(\boldsymbol{\theta}) = 0.5$. The outer and inner black dashed lines indicate the boundaries of the response regimes corresponding to $\beta(\boldsymbol{\theta}) = 0.325$ and $\beta(\boldsymbol{\theta}) = 0.675$, respectively.

Despite its effectiveness, DR$^2$EI is subject to certain limitations. For instance, the method requires the user to select a set of sensitivity parameters $\Delta$ prior to its application. Additionally, the method's efficiency is dependent on the quality and effectiveness of the active sequential sampling approach based on Gaussian Process Regression. It is also important to note that DR$^2$EI is not guaranteed to identify all of the dynamical responses of a given system; rather, it can only identify those responses associated with hypervolumes in the parameter space that are sufficiently large.

Nevertheless, the active sequential sampling algorithm, or even dense enough random sampling, statistically provides a means to assume that the vast majority, if not all, statistically significant regimes will be captured and classified. This is precisely the goal of the UQ-based sequential sampling methodology proposed in this study.

Future studies will aim to reduce the dependence on the choice of predefined parameters by implementing an adaptive selection of those parameters. This will be achieved through an iterative scan of the parameter space, enabling the method to effectively identify the response regimes of the system without requiring prior knowledge of the system's behavior. Additionally, comparing different sampling methodologies that focus on the derivative of the behavior function $\beta'(\boldsymbol{\theta})$ will be explored to further refine the boundaries in the parameter space and improve exploration efficiency. These efforts will contribute to the development of an even more effective and robust version of DR$^2$EI that will enable comprehensive and efficient exploration of complex dynamical systems.

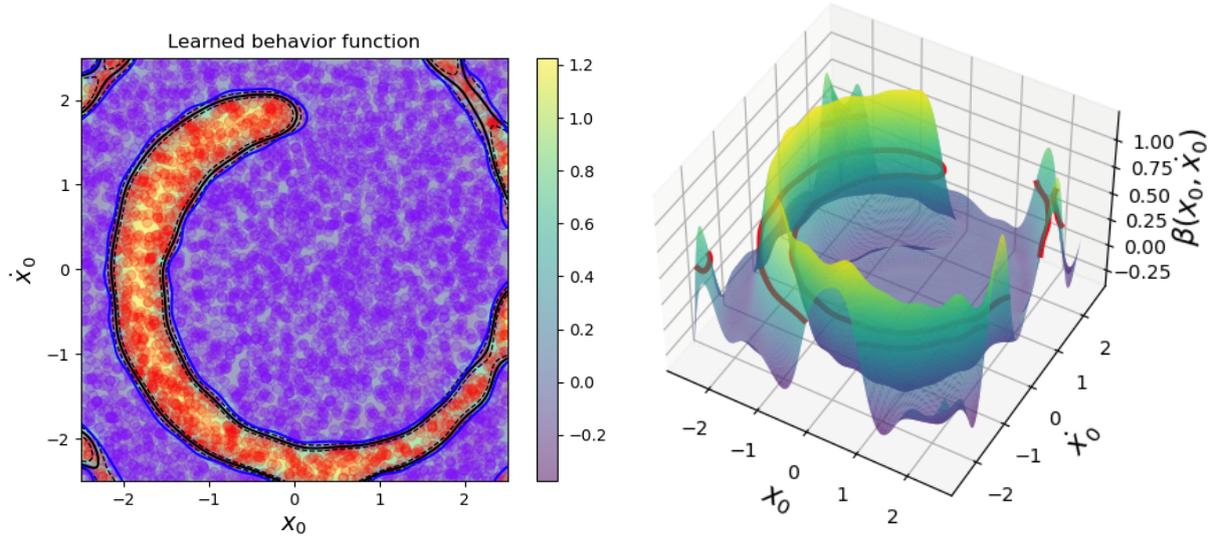

Figure 10: Left- Performance evaluation of DR$^2$EI method over the parameter (ICs) space of the Duffing oscillator with parameters $\bar{\epsilon} = 0.01, F = 2, \bar{\alpha} = 6$ and $\bar{\lambda} = 0.5$. The learned behavior function, which corresponds to the mean function of the GPR, is used to identify the classified sampled points from the parameter space. Red and purple dots indicate the sample parameters corresponding to the high- and low-amplitude periodic regimes, respectively. The solid black line in the figure represents the estimated boundary of the response regimes identified by the DR$^2$EI method, which corresponds to $\beta(\boldsymbol{\theta}) = 0.5$. The outer and inner black dashed lines correspond to $\beta(\boldsymbol{\theta}) = 0.325$ and $\beta(\boldsymbol{\theta}) = 0.675$, respectively. The solid blue line corresponds to the computed steady-state response amplitudes obtained through direct numerical integration of the system's equations of motion. Right- 3D view of the learned behavior function $\beta(\boldsymbol{\theta})$, and the separation contiur (red solid line) that corresponds to $\beta(\boldsymbol{\theta}) = 0.5$.

## Declarations

**Conflict of Interest** The author declares that he has no conflict of interest.

**Availability of data and material** The data that supports the findings of this study is available from the author upon request.

# List of Abbreviations

| | |
|---|---|
| CDF | Cumulative Distribution Function |
| DBSCAN | density-based spatial clustering of applications with noise |
| DOF | Degree of Freedom |
| DR$^2$EI | Data-Driven Response Regimes Exploration and Identification |
| EI | Expected Improvement |
| EOM | Equation of motion |
| FFT | Fast Fourier Transform |
| GP | Gaussian Process |
| GPR | Gaussian Process Regression |
| IC | Initial Condition |
| MAP | Maximum *a-posteriori* |
| ML | Machine learning |
| PCA | Principal Component Analysis |
| PDF | Probability Density Function |
| RBF | Radial Basis Function |
| SVD | Singular Value Decomposition |
| t-SNE | t-Distributed Stochastic Neighbor Embedding |
| UQ | Uncertainty Quantification |

# List of Symbols

| | |
|---|---|
| $\bar{\omega}, \bar{\lambda}, \bar{\epsilon}, \bar{\alpha}, \bar{F}$ | Natural frequency, damping coefficient, small parameter, cubic nonlinearity coefficient, and forcing amplitude of the Duffing Equation |
| $\beta(\boldsymbol{\theta})$ | Latent behavior function that maps parameter sets to their corresponding response regime index |
| $\beta_{max}$ | The current best estimate of the latent behavior function $\beta(\boldsymbol{\theta})$ value for $\boldsymbol{\theta}$ |
| $\boldsymbol{\sigma}(\boldsymbol{\theta}, \mathcal{D})$ | The standard deviation of the GPR |
| $\boldsymbol{\theta} \in \Theta \in \mathbb{R}^n$ | Parameter vector sampled from the parameter space |
| $\boldsymbol{\gamma} \in \boldsymbol{\Gamma}$ | Hyper-parameters vector of the GPR, $\boldsymbol{\gamma} = \{\sigma_n, l, \sigma_l\}$ and its underlying hyper-parameter space |
| $\boldsymbol{v}(\boldsymbol{\theta}) \in \mathbb{R}^m$ | The embedding vector that corresponds to parameter set $\boldsymbol{\theta}$, $m \ll n_t$ |
| $\Delta$ | The set of sensitivity parameters that are chosen by the user a-priori for the DR$^2$EI algorithm |
| $\dot{()}$ | Derivation with respect to time $t$ |
| $\epsilon \sim \mathcal{N}(0, \sigma_n^2)$ | Zero-mean Gaussian noise with standard deviation of $\sigma_n$ |
| $\hat{\epsilon}$, MinPts | The radius of the neighborhood around each point in the embedding space, and the minimum number of points required to form a dense region or cluster, respectively |
| $\mathbb{E}$ | Expected value |
| $\mathcal{D} = \{\boldsymbol{\theta}_i, \tau_i\}, \boldsymbol{\theta}_i \in \tilde{\Theta}$ | Set of parameter sets and their corresponding estimated value of the behavior function |
| $\mu, \Sigma, \sigma$ | Mean, variance, and standard deviation of the posterior GP |
| $\mu_0, \Sigma_0, K$ | Mean, variance, and covariance matrix of the prior GP |
| $\omega$ | Angular frequency |
| $\partial \tilde{\Theta} \in \mathbb{R}^{n-1}$ | The circumference of the hypervolume $\tilde{\Theta}$ in parameter space $\Theta$ |
| $\Phi(Z), \phi(Z)$ | cumulative distribution function (CDF) and probability density function (PDF) of the standard normal distribution, respectively |
| $\Psi \in \mathbb{R}^m$ | Embedding space |
| $\tau$ | Noisy observation |
| $\Theta, \tilde{\Theta} \in \mathbb{R}^n$ | Parameter space and set of sampled parameter sets from the parameter space, respectively |
| $\Theta_0 \in \mathbb{R}^n$ | Initial randomly selected parameter set |
| $\zeta$ | Parameter that controls the trade-off between exploration and exploitation |
| $C_i, u_i$ | The $i$-th cluster containing all data points that are directly or indirectly reachable from point $u_i$ |
| $d(u_i, u_j)$ | The distance between data points $u_i$ and $u_j$ in the embedding space |
| $g : \Theta \to \Psi$ | Transformation function that projects a given time series to a lower order embedding space |
| $k$ | Frequency index |
| $N$ | Number of degrees of freedom |
| $N_f$ | Number of frequency bins considered in the FFT |
| $N_r$ | The number of distinct clusters identified by the unsupervised clustering algorithm and is also the number of response regimes identified by the DR$^2$EI method for a given dynamical system |
| $s, b, \rho$ | Parameters on the Lorenz system |
| $t \in \mathbb{R}^{n_t}$ | Time variable |
| $t_f$ | Termination time of the time series |
| $T_{max}$ | Maximum number of iterations of the sequential sampling algorithm |
| $u(t|\boldsymbol{\theta}) \in \mathbb{R}^{n_t}$ | Time series that corresponds to parameter set $\boldsymbol{\theta}$ |
| $Z$ | standard normal random variable |

# References

[1] A. H. Nayfeh, D. T. Mook, Nonlinear Oscillations, Vol. 27, John Wiley and Sons, 2008.

[2] S. H. Strogatz, Nonlinear Dynamics and Chaos: With Applications to Physics, Biology, Chemistry, and Engineering, CRC press, 2018.

[3] H. Allaka, M. Farid, M. Groper, Mitigation of vertical motion in planing crafts for enhanced operationability in seaways using passive energy absorbers – a test of concept, Ocean Engineering 264 (2022) 112434.

[4] M. Farid, O. Gendelman, Response regimes in equivalent mechanical model of moderately nonlinear liquid sloshing, Nonlinear Dynamics 92 (4) (2018) 1517–1538.

[5] M. Farid, N. Levy, O. Gendelman, Vibration mitigation in partially liquid-filled vessel using passive energy absorbers, Journal of Sound and Vibration 406 (2017) 51–73.

[6] M. Farid, O. Gendelman, Internal resonances and dynamic responses in equivalent mechanical model of partially liquid-filled vessel, Journal of Sound and Vibration 379 (2016) 191–212.

[7] M. Farid, O. Gendelman, Response regimes in equivalent mechanical model of strongly nonlinear liquid sloshing, International Journal of Non-Linear Mechanics 94 (2017) 146–159, a Conspectus of Nonlinear Mechanics: A Tribute to the Oeuvres of Professors G. Rega and F. Vestroni.

[8] M. Farid, O. Gendelman, Internal resonances and dynamic responses in equivalent mechanical model of partially liquid-filled vessel, Procedia Engineering 199 (2017) 3440–3443, x International Conference on Structural Dynamics, EURODYN 2017.

[9] L. M. Pecora, T. L. Carroll, Synchronization in chaotic systems, Physical Review Letters 64 (8) (1990) 821.

[10] H. Kantz, T. Schreiber, Nonlinear Time Series Analysis, Vol. 7, Cambridge University Press, 2004.

[11] E. Ott, C. Grebogi, J. A. Yorke, Controlling chaos, Physical review letters 64 (11) (1990) 1196.

[12] S. Boccaletti, V. Latora, Y. Moreno, M. Chavez, D.-U. Hwang, Complex networks: Structure and dynamics, Physics reports 424 (4-5) (2006) 175–308.

[13] I. Syarif, A. Prugel-Bennett, G. Wills, Svm parameter optimization using grid search and genetic algorithm to improve classification performance, TELKOMNIKA (Telecommunication Computing Electronics and Control) 14 (4) (2016) 1502–1509.

[14] S. L. Marple Jr, W. M. Carey, Digital spectral analysis with applications (1989).

[15] P. D. Welch, The use of the fast fourier transform for the estimation of power spectra: A method based on time averaging over short, modified periodograms, IEEE Transactions on Audio and Electroacoustics 15 (2) (1967) 70–73.

[16] W. A. Gardner, Exploitation of spectral redundancy in cyclostationary signals, IEEE Signal processing magazine 8 (2) (1991) 14–36.

[17] E. O. Brigham, The fast Fourier transform and its applications, Prentice-Hall, Inc., 1988.

[18] S. Mallat, A wavelet tour of signal processing, Elsevier, 1999.

[19] J. Guckenheimer, P. Holmes, Nonlinear oscillations, dynamical systems, and bifurcations of vector fields, Vol. 42, Springer Science and Business Media, 2013.

[20] A. Krause, C. Ong, Contextual gaussian process bandit optimization, Advances in neural information processing systems 24 (2011).

[21] N. Srinivas, A. Krause, S. M. Kakade, M. Seeger, Gaussian process optimization in the bandit setting: No regret and experimental design, arXiv preprint arXiv:0912.3995 (2009).

[22] C. K. Williams, C. E. Rasmussen, Gaussian processes for machine learning, Vol. 2, MIT press Cambridge, MA, 2006.

[23] C. K. Williams, Prediction with gaussian processes: From linear regression to linear prediction and beyond, Learning in graphical models (1998) 599–621.

[24] D. J. MacKay, et al., Introduction to gaussian processes, NATO ASI series F computer and systems sciences 168 (1998) 133–166.

[25] E. Snelson, Z. Ghahramani, Sparse gaussian processes using pseudo-inputs, Advances in neural information processing systems 18 (2005).

[26] D. P. Kingma, M. Welling, Auto-encoding variational bayes, arXiv preprint arXiv:1312.6114 (2013).

[27] I. Higgins, L. Matthey, A. Pal, C. Burgess, X. Glorot, M. Botvinick, S. Mohamed, A. Lerchner, beta-vae: Learning basic visual concepts with a constrained variational framework, International Conference on Learning Representations (2017).

[28] Y. Liang, H. Lee, S. Lim, W. Lin, K. Lee, C. Wu, Proper orthogonal decomposition and its applications—part i: Theory, Journal of Sound and vibration 252 (3) (2002) 527–544.

[29] P. J. Schmid, Dynamic mode decomposition of numerical and experimental data, Journal of fluid mechanics 656 (2010) 5–28.

[30] M. Ghil, M. Allen, M. Dettinger, K. Ide, D. Kondrashov, M. Mann, A. W. Robertson, A. Saunders, Y. Tian, F. Varadi, et al., Advanced spectral methods for climatic time series, Reviews of geophysics 40 (1) (2002) 3–1.

[31] M. Ester, H.-P. Kriegel, J. Sander, X. Xu, A density-based algorithm for discovering clusters in large spatial databases with noise, Proceedings of 2nd International Conference on Knowledge Discovery and Data Mining (1996).

[32] H. Hotelling, Analysis of a complex of statistical variables into principal components., Journal of educational psychology 24 (6) (1933) 417.

[33] L. Van der Maaten, G. Hinton, Visualizing data using t-sne., Journal of machine learning research 9 (11) (2008).